\documentclass[showpacs,preprintnumbers,amsmath,amssymb,nofootinbib,floatfix]{revtex4}
\usepackage[dvips]{graphicx}% Include figure files
\usepackage{dcolumn}% Align table columns on decimal point
\usepackage{bm}% bold math

\def\mapgeq{\mathbin{\lower.3ex\hbox{$\buildrel>\over{\smash{\scriptstyle\sim}\vphantom{_x}}$}}}
\def\mapleq{\mathbin{\lower.3ex\hbox{$\buildrel<\over{\smash{\scriptstyle\sim}\vphantom{_x}}$}}}
\def\mapgeqeq{\mathbi{\lower.3ex\hbox{$\buildrel>\over{\smash{\scriptstyle\approx}\vphantom{_2}}$}}}
\def\mapleqeq{\mathbin{\lower.3ex\hbox{$\buildrel<\over{\smash{\scriptstyle\approx}\vphantom{_2}}$}}}
\def\Journal#1#2#3#4{{#1} {\bf #2} (#4) #3}
\def\MPL{Mod. Phys. Lett. A}

\def\NPB{Nucl. Phys. B}

\def\NPSUPPL{Nucl. Phys. Proc. Suppl.}
\def\PLB{{Phys. Lett.} B}

\def\PLBOLD{Phys. Lett.}
\def\PRL{Phys. Rev. Lett.}
\def\RMP{Rev. Mod. Phys.}
\def\PRD{Phys. Rev. D}

\def\PTP{Prog. Theor. Phys.}
\def\JHEP{JHEP}

\def\EPJ{Euro. Phys. J. C}

\def\JETPUSSR{Sov. Phys. JETP}
\def\ZETP{Zh. Eksp. Teor. Piz.}

\def\IJMP{Int. J. Mod. Phys. A}

\def\JPG{J. Phys. G}

\def\NIMA{Nucl. Instrum. Methods A}
\def\NJP{New. J. Phys}
\def\Erratum{Erratum-ibid}

\begin{document}

\preprint{TOKAI-HEP/TH-0503}

\title{$\mu$-$\tau$ Symmetry and Maximal CP Violation}
% Force line breaks with \\

\author{Teruyuki Kitabayashi}
\email{teruyuki@keyaki.cc.u-tokai.ac.jp}

% \affiliation[Also at ]{Physics Department, XYZ University.}%Lines break automatically or can be forced with \\
\author{Masaki Yasu\`{e}}%
\email{yasue@keyaki.cc.u-tokai.ac.jp}
\affiliation{\vspace{5mm}%
\sl Department of Physics, Tokai University,\\
1117 Kitakaname, Hiratsuka, Kanagawa 259-1292, Japan\\
}

\date{April, 2005}% It is always \today, today,

%%-------------------------------------------------
%% Abstruct
%%-------------------------------------------------
\begin{abstract}
We argue the possibility that a real part of a flavor neutrino mass matrix only respects a $\mu$-$\tau$ symmetry.  This possibility is shown to be extended to more general case with a phase parameter $\theta$, where the $\mu$-$\tau$ symmetric part has a phase of $\theta/2$.  This texture shows maximal CP violation and no Majorana CP violation.
\end{abstract}

%\keywords{Keywords: neutrino mass, radiative mechanism, lepton triplet}%Use showkeys class option if keyword
\maketitle
%%-------------------------------------------------
%% Main body
%%-------------------------------------------------
The present experimental data on neutrino oscillations \cite{SK,experiments} indicate the mixing angles \cite{NeutrinoSummary} satisfying
%%%%%%%%%%%%%%%%%%%%
\begin{eqnarray}
&&0.70 < \sin^2 2\theta_\odot < 0.95,\quad
0.92 < \sin^22\theta_{atm},\quad
\sin^2\theta_{CHOOZ} <0.05,
\label{Eq:MixingData}
\end{eqnarray}
%%%%%%%%%%%%%%%%%%%%
where $\theta_\odot$ is the solar neutrino mixing angle, $\theta_{atm}$ is the atmospheric neutrino mixing angle and $\theta_{CHOOZ}$ is for the mixing angle between $\nu_e$ and $\nu_\tau$.  These mixing angles are identified with the mixings among three flavor neutrinos, $\nu_e$, $\nu_\mu$ and $\nu_\tau$, yielding three massive neutrinos, $\nu_{1,2,3}$: $\theta_{12}=\theta_\odot$, $\theta_{23}=\theta_{atm}$ and $\theta_{13}=\theta_{CHOOZ}$.  These data seem to be consistent with the presence of a $\mu$-$\tau$ symmetry \cite{Nishiura,mu-tau,mu-tau1,mu-tau2} in the neutrino sector, which provides maximal atmospheric neutrino mixing with $\sin^22\theta_{23}=1$ as well as $\sin\theta_{13}=0$.

Although neutrinos gradually reveal their properties in various experiments since the historical Super-Kamiokande confirmation of neutrino oscillations \cite{SK}, we expect to find yet unknown property related to CP violation \cite{CPViolation}. The effect of the presence of a leptonic CP violation can be described by four phases in the PMNS neutrino mixing matrix, $U_{PMNS}$ \cite{PMNS}, to be denoted by one Dirac phase of $\delta$ and three Majorana phases of $\beta_{1,2,3}$ as $U_{PMNS}=U_\nu K$ \cite{CPphases} with
%%%%%%%%%%%%%%%%%%%%
\begin{eqnarray}
U_\nu&=&\left( \begin{array}{ccc}
  c_{12}c_{13} &  s_{12}c_{13}&  s_{13}e^{-i\delta}\\
  -c_{23}s_{12}-s_{23}c_{12}s_{13}e^{i\delta}
                                 &  c_{23}c_{12}-s_{23}s_{12}s_{13}e^{i\delta}
                                 &  s_{23}c_{13}\\
  s_{23}s_{12}-c_{23}c_{12}s_{13}e^{i\delta}
                                 &  -s_{23}c_{12}-c_{23}s_{12}s_{13}e^{i\delta}
                                 & c_{23}c_{13},\\
\end{array} \right),
\nonumber \\
K &=& {\rm diag}(e^{i\beta_1}, e^{i\beta_2}, e^{i\beta_3}),
\label{Eq:U_nu}
\end{eqnarray}
%%%%%%%%%%%%%%%%%%%%
where $c_{ij}=\cos\theta_{ij}$ and $s_{ij}=\sin\theta_{ij}$ ($i,j$=1,2,3) and Majorana CP violation is specified by two combinations made of $\beta_{1,2,3}$ such as $\beta_i-\beta_3$ in place of $\beta_i$ in $K$.  To examine such effects of CP violation, there have been various discussions \cite{CPinMixing} including those on the possible textures of flavor neutrino masses \cite{MassTextureCP1,MassTextureCP2,CP2-3,CP2-3-1}. 

In this note, we would like to focus on the role of the $\mu$-$\tau$ symmetry in models with CP violation \cite{CP2-3,CP2-3-1}, which can be implemented by introducing complex flavor neutrino masses.  The $\mu$-$\tau$ symmetric texture gives $\sin\theta_{13}=0$ as well as maximal atmospheric neutrino mixing characterized by $c_{23}=\sigma s_{23}=1/\sqrt{2}$ ($\sigma=\pm 1$).  Because of $\sin\theta_{13}=0$, Dirac CP violation is absent in Eq.(\ref{Eq:U_nu}) and CP violation becomes of the Majorana type.  Since the $\mu$-$\tau$ symmetry is expected to be approximately realized, its breakdown is signaled by $\sin\theta_{13} \neq 0$.  To have $\sin\theta_{13} \neq 0$, we discuss another implementation of the $\mu$-$\tau$ symmetry such that the symmetry is only respected by the real part of $M_\nu$.  The discussion is based on more general case, where $M_\nu$ is controlled by one phase to be denoted by $\theta$ and the specific value of $\theta=0$ yields the $\mu$-$\tau$ symmetric real part.  It turns out that Majorana CP violation is absent because all three Majorana phases are calculated to be $-\theta/4$ while Dirac CP violation becomes maximal.

Our complex flavor neutrino mass matrix of $M_\nu$ is parameterized by
%%%%%%%%%%%%%%%%%%%%
\begin{eqnarray}
&& M_\nu = \left( {\begin{array}{*{20}c}
	M_{ee} & M_{e\mu} & M_{e\tau}  \\
	M_{e\mu} & M_{\mu\mu} & M_{\mu\tau}  \\
	M_{e\tau} & M_{\mu\tau} & M_{\tau\tau}  \\
\end{array}} \right),
\label{Eq:NuMatrixEntries}
\end{eqnarray}
%%%%%%%%%%%%%%%%%%%%
where $U^T_{PMNS}M_\nu U_{PMNS}$=diag.($m_1$, $m_2$, $m_3$).\footnote{It is understood that the charged leptons and neutrinos are rotated, if necessary, to give diagonal charged-current interactions and to define the flavor neutrinos of $\nu_e$, $\nu_\mu$ and $\nu_\tau$.}  The mixing angles have been calculated in the Appendix of Ref.\cite{MaxCPmass} and satisfy
%%%%%%%%%%%%%%%%%%%%
\begin{eqnarray}
&&
{\sin 2\theta _{12} \left( {\lambda _1  - \lambda _2 } \right) + 2\cos 2\theta _{12} X = 0},
\label{Eq:theta12}\\
&&
{\sin 2\theta _{13} \left( {M_{ee}e^{ - i\delta }  - \lambda _3 e^{i\delta } } \right) + 2\cos 2\theta _{13} Y} = 0,
\label{Eq:theta13}\\
&&
\left( M_{\tau\tau} - M_{\mu\mu}\right)\sin 2\theta_{23}  - 2 M_{\mu\tau}\cos 2\theta_{23}= 2s_{13} e^{ - i\delta } X,
\label{Eq:theta23}
\end{eqnarray}
%%%%%%%%%%%%%%%%%%%%
and neutrino masses are given by
%%%%%%%%%%%%%%%%%%%%
\begin{eqnarray}
&&
m_1 e^{ - 2i\beta_1 }  = \frac{{\lambda_1  + \lambda_2 }}{2} - \frac{X}{{\sin 2\theta_{12} }},
\quad
m_2 e^{ - 2i\beta_2 }  = \frac{{\lambda_1  + \lambda_2 }}{2} + \frac{X}{{\sin 2\theta_{12} }},
\nonumber\\
&&
m_3 e^{ - 2i\beta_3 }  =\frac{c_{13}^2 \lambda _3 - s_{13}^2 e^{-2i\delta } M_{ee} }{\cos 2\theta _{13} },
\label{Eq:masses_1-2-3}
\end{eqnarray}
%%%%%%%%%%%%%%%%%%%%
where
%%%%%%%%%%%%%%%%%%%%
\begin{eqnarray}
&&
\lambda_1  = c_{13}^2 M_{ee} - 2c_{13} s_{13} e^{i\delta } Y + s_{13}^2 e^{2i\delta }\lambda_3,
\quad
\lambda_2  = c_{23}^2 M_{\mu\mu} + s_{23}^2 M_{\tau\tau} - 2s_{23} c_{23} M_{\mu\tau},
\nonumber\\
&&
\lambda_3  = s_{23}^2 M_{\mu\mu} + c_{23}^2 M_{\tau\tau} + 2s_{23} c_{23} M_{\mu\tau},
\label{Eq:Parameters}\\
&&
X = \frac{c_{23} M_{e\mu} - s_{23} M_{e\tau}}{c_{13}},
\quad
Y = s_{23} M_{e\mu} + c_{23} M_{e\tau}.
\label{Eq:X-Y}
\end{eqnarray}
%%%%%%%%%%%%%%%%%%%%

To clarify the importance of the $\mu$-$\tau$ symmetry, which accommodates maximal atmospheric neutrino mixing and $\sin\theta_{13}=0$, we first review what conditions are imposed by the requirement of $\sin\theta_{13}=0$.  From Eq.(\ref{Eq:theta13}), we require that
%%%%%%%%%%%%%%%%%%%%
\begin{eqnarray}
&& 
Y=s_{23}M_{e\mu}+c_{23}M_{e\tau}=0,
\label{Eq:Y=0}
\end{eqnarray}
%%%%%%%%%%%%%%%%%%%%
giving rise to $\tan\theta_{23}=-{\rm Re}(M_{e\tau})/{\rm Re}(M_{e\mu})=-{\rm Im}(M_{e\tau})/{\rm Im}(M_{e\mu})$.  Since $\sin\theta_{13}=0$, Eq.(\ref{Eq:theta23}) reads
%%%%%%%%%%%%%%%%%%%%
\begin{eqnarray}
&& 
\left(M_{\tau\tau} - M_{\mu\mu}\right)\sin 2\theta_{23} = 2M_{\mu\tau}\cos 2\theta_{23}.
\label{Eq:Delta=0-1}
\end{eqnarray}
%%%%%%%%%%%%%%%%%%%%
These are the well known relations that determine $\theta_{23}$ if $\sin\theta_{13}=0$. Maximal atmospheric neutrino mixing arises if 
%%%%%%%%%%%%%%%%%%%%
\begin{eqnarray}
&& 
M_{\tau\tau}=M_{\mu\mu},
\label{Eq:mu-tau-maximal-1}
\end{eqnarray}
%%%%%%%%%%%%%%%%%%%%
 which in turn gives
%%%%%%%%%%%%%%%%%%%%
\begin{eqnarray}
&& 
M_{e\tau} = -\sigma M_{e\mu}.
\label{Eq:mu-tau-maximal-2}
\end{eqnarray}
%%%%%%%%%%%%%%%%%%%%
The flavor neutrino masses satisfying Eqs.(\ref{Eq:mu-tau-maximal-1}) and (\ref{Eq:mu-tau-maximal-2}) suggest the presence of $\mu$-$\tau$ symmetry in neutrino physics.  The remaining mixing angle of $\theta_{12}$ satisfies
%%%%%%%%%%%%%%%%%%%%
\begin{eqnarray}
&& 
M_{\mu\mu} - \sigma M_{\mu\tau} = M_{ee} + \frac{{2\sqrt 2 }}{{\tan 2\theta _{12} }}M_{e\mu},
\label{Eq:tan12-single}
\end{eqnarray}
%%%%%%%%%%%%%%%%%%%%
which determines the definite correlation of the phases of the flavor neutrino masses.  

In place of Eqs.(\ref{Eq:Y=0}) and (\ref{Eq:Delta=0-1}), using  a Hermitian matrix of ${\rm\bf M}=M^\dagger_\nu M_\nu$, we can find that $\tan\theta_{23}=-{\rm Re}\left({\rm\bf M}_{e\tau}\right)/{\rm Re}\left({\rm\bf M}_{e\mu}\right)=-{\rm Im}\left({\rm\bf M}_{e\tau}\right)/{\rm Im}\left({\rm\bf M}_{e\mu}\right)$, where ${\rm\bf M}_{e\mu} = M^\ast_{ee}M_{e\mu}+M^\ast_{e\mu}M_{\mu\mu}+M^\ast_{e\tau}M_{\mu\tau}$ and ${\rm\bf M}_{e\tau} = M^\ast_{ee}M_{e\tau}+M^\ast_{e\mu}M_{\mu\tau}+M^\ast_{e\tau}M_{\tau\tau}$. In addition, we have argued that $\tan\theta_{23}$ is directly determined by $\tan \theta _{23}  = {\rm Im}( {\rm\bf M}_{e\mu})/{\rm Im}({\rm\bf M}_{e\tau})$ satisfied in any models with complex neutrino masses irrespective of the values of $\sin\theta_{13}$ \cite{GeneralCP}. Both expressions of $\tan\theta_{23}$ are compatible if $({\rm Im}({\rm\bf M}_{e\mu}))^2+({\rm Im}({\rm\bf M}_{e\tau}))^2$=0, yielding ${\rm Im}({\rm\bf M}_{e\mu})={\rm Im}({\rm\bf M}_{e\tau})=0$.  Since the Dirac CP violation phase is absent for $\sin\theta_{13}=0$, ${\rm\bf M}$ with the Majorana phases cancelled is necessarily real.  In fact, we obtain that ${\rm\bf M}_{e\mu} = c_{12} s_{12} c_{23} (m_2^2  - m_1^2)$ and ${\rm\bf M}_{e\tau}=-\tan\theta_{23}{\rm\bf M}_{e\mu}$ which automatically satisfy  ${\rm Im}({\rm\bf M}_{e\mu})={\rm Im}({\rm\bf M}_{e\tau})=0$.

We next argue the implementation of the $\mu$-$\tau$ symmetry based on the observation that it is sufficient for the symmetry to be respected by the real part of $M_\nu$.  From the discussions developed in Ref.\cite{MaxCPmass}, it can be extended to more general case, where the real and imaginary parts are, respectively, replaced by $(z + e^{i\theta}z^\ast)/2$ $(\equiv z_+)$ and $(z - e^{i\theta}z^\ast)/2$ $(\equiv z_-)$ for a complex number of $z$ and the phase parameter of $\theta$.  It is useful to notice that $z_+ = e^{i\theta/2}{\rm Re}(e^{-i\theta/2}z)$ and $z_- = ie^{i\theta/2}{\rm Im}(e^{-i\theta/2}z)$. The relevant mass matrix is provided by one of the textures found in Ref.\cite{MaxCPmass}:
%%%%%%%%%%%%%%%%%%%%
\begin{eqnarray}
&& 
	M_\nu = \left( {\begin{array}{*{20}c}
   M_{ee} & M_{e\mu} & -\sigma e^{i\theta}M^\ast_{e\mu} \\
   M_{e\mu} & M_{\mu\mu} & M_{\mu\tau}  \\
   -\sigma e^{i\theta}M^\ast_{e\mu} & M_{\mu\tau} & e^{i\theta}M^\ast_{\mu\mu}  \\
\end{array}} \right),
\label{Eq:Mnu-1}
\end{eqnarray}
%%%%%%%%%%%%%%%%%%%%
where $M_{ee,\mu\tau}=e^{i\theta}M^\ast_{ee,\mu\tau}$, equivalently $(M_{ee,\mu\tau})_-=0$, is imposed.  This texture gives
%%%%%%%%%%%%%%%%%%%%
\begin{eqnarray}
&& 
\tan 2\theta _{12}  =
2\sqrt 2 \frac{\cos 2\theta _{13} \left( M_{e\mu} \right)_+}{c_{13} \left[ \left( 1-3s^2_{13}\right) \left( M_{\mu\mu}\right)_+ - c_{13}^2 \left(\sigma\left(M_{\mu\tau}\right)_++ \left(M_{ee}\right)_+ \right) \right]},
\label{Eq:Angle12-1}\\
&&
\tan 2\theta _{13} e^{i\delta } = 2\sqrt 2 \frac{\sigma \left( M_{e\mu} \right)_-}{\left( M_{\mu\mu} \right)_+ + \sigma\left(  M_{\mu\tau}\right)_+ + \left( M_{ee}\right)_+}.
\label{Eq:Angle13-1}
\end{eqnarray}
%%%%%%%%%%%%%%%%%%%%
As discussed in Ref.\cite{MaxCPmass}, these expressions yield real values of $\tan 2\theta_{12,13}$ because of the property that ${z^\prime_+}/{z_+}={\rm Re}(e^{-i\theta/2}z^\prime)/{\rm Re}(e^{-i\theta/2}z)$ and ${z ^\prime_-}/{z _+}=i{\rm Im}(e^{-i\theta/2}z^\prime)/{\rm Re}(e^{-i\theta/2}z)$ for any complex values of $z$ and $z^\prime$. As a result, $\delta=\pm \pi/2$ is derived and $M_\nu$ gives maximal CP violation.

A texture with the Dirac CP violation phase related to the $\mu$-$\tau$ symmetric texture is obtained by decomposing $z$ and $e^{i\theta}z^\ast$  into $z_+$ and $z_-$ and turns out to be $M_\nu = M_{+\nu}+M_{-\nu}$ with
%%%%%%%%%%%%%%%%%%%%
\begin{eqnarray}
&& 
	M_{+\nu} = \left( {\begin{array}{*{20}c}
   (M_{ee})_+ & (M_{e\mu})_+ & -\sigma (M_{e\mu})_+ \\
   (M_{e\mu})_+ & (M_{\mu\mu})_+ & (M_{\mu\tau})_+  \\
   -\sigma (M_{e\mu})_+ & (M_{\mu\tau})_+ & (M_{\mu\mu})_+  \\
\end{array}} \right) = e^{i\theta/2}{\rm Re}\left(e^{-i\theta/2} M_\nu\right),
\nonumber\\
&&
	M_{-\nu} = \left( {\begin{array}{*{20}c}
   0 & (M_{e\mu})_- & \sigma (M_{e\mu})_- \\
   (M_{e\mu})_- & (M_{\mu\mu})_- & 0  \\
   \sigma (M_{e\mu})_- & 0 & -(M_{\mu\mu})_-  \\
\end{array}} \right) = ie^{i\theta/2}{\rm Im}\left(e^{-i\theta/2} M_\nu\right),
\label{Eq:Texture23-complex}
\end{eqnarray}
%%%%%%%%%%%%%%%%%%%%
which shows that $M_{+\nu}$ has a phase $\theta/2$ modulo $\pi$ while $M_{-\nu}$ has a phase $(\theta+\pi)/2$ modulo $\pi$.  The $\mu$-$\tau$ symmetry exists in $M_{+\nu}$ because Eqs.(\ref{Eq:mu-tau-maximal-1}) and (\ref{Eq:mu-tau-maximal-2}) are satisfied but is explicitly broken by $M_{-\nu}$. Therefore, this texture shows ``incomplete" $\mu$-$\tau$ symmetry \cite{CP2-3-1}.  Since $M_{+\nu}$ does not contribute to $\sin\theta_{13}$, $\sin\theta_{13}$ should be proportional to the flavor neutrino masses in $M_{-\nu}$.  In fact, it is proportional to $( M_{e\mu})_-$ in Eq.(\ref{Eq:Angle13-1}).  To speak of the Majorana phases, we have to determine neutrino masses, which can be computed from Eq.(\ref{Eq:masses_1-2-3}) and are given by
%%%%%%%%%%%%%%%%%%%%
\begin{eqnarray}
&& 
m_1 e^{ - 2i\beta _1 }  = 
\left( M_{\mu\mu} \right)_+-\sigma \left( M_{\mu\tau} \right)_+- \frac{1 + \cos 2\theta _{12} }{\sin 2\theta _{12} }\frac{\sqrt 2 \left( M_{e\mu} \right)_+}{c_{13} },
\nonumber\\
&& 
m_2 e^{ - 2i\beta _2 }  = 
\left( M_{\mu\mu} \right)_+-\sigma \left( M_{\mu\tau} \right)_+ + \frac{1 - \cos 2\theta _{12} }{\sin 2\theta _{12} }\frac{\sqrt 2 \left( M_{e\mu} \right)_+}{c_{13} },
\nonumber\\
&& 
m_3 e^{ - 2i\beta _3 }  = \frac{c^2_{13}\left( \left( M_{\mu\mu} \right)_+ +\sigma \left( M_{\mu\tau} \right)_+\right)+ s^2_{13}\left( M_{ee} \right)_+}{\cos 2\theta _{13}}.
\label{Eq:Masses-1}
\end{eqnarray}
%%%%%%%%%%%%%%%%%%%%
Since $z_+ = e^{i\theta/2}{\rm Re}(e^{-i\theta/2}z)$, the texture gives three Majorana phases calculated to be: $\beta_{1,2,3}=-\theta/4$ modulo $\pi/2$.  The common phase does not induce Majorana CP violation.  This result reflects the fact that the source of the Majorana phases is the phase of $M_\nu$ in Eq.(\ref{Eq:Texture23-complex}) equal to $\theta/2$, which can be rotated away by redefining appropriate fields. The remaining imaginary part ${\rm Im}(e^{-i\theta/2} M_\nu)$ supplies the Dirac phase $\delta$.  Therefore, our proposed mass matrix becomes ${\rm Re}(e^{-i\theta/2} M_\nu)+i{\rm Im}(e^{-i\theta/2} M_\nu)$, which is equivalent to $M_\nu$ with $\theta$=0. No CP violating Majorana phases exist in our mass matrix.

The simplest choice of $\theta=0$ provides the case where the real part of $M_\nu$ respects the $\mu$-$\tau$ symmetry. This texture has been discussed in Ref.\cite{MassTextureCP2,GeneralCP}, which takes the form of
%%%%%%%%%%%%%%%%%%%%
\begin{eqnarray}
&& 
	M^{\mu-\tau}_{\nu} = {\rm Re}\left( {\begin{array}{*{20}c}
   M_{ee} & M_{e\mu} & -\sigma M_{e\mu} \\
   M_{e\mu} & M_{\mu\mu} & M_{\mu\tau}  \\
   -\sigma M_{e\mu} & M_{\mu\tau} & M_{\mu\mu}  \\
\end{array}} \right)
	+ i {\rm Im}\left( {\begin{array}{*{20}c}
   0 & M_{e\mu} & \sigma M_{e\mu} \\
   M_{e\mu} & M_{\mu\mu} & 0  \\
   \sigma M_{e\mu} & 0 & -M_{\mu\mu} \\
\end{array}} \right),
\label{Eq:Texture23-direct}
\end{eqnarray}
%%%%%%%%%%%%%%%%%%%%
where the real part is the well-known $\mu$-$\tau$ symmetric texture as expected while the imaginary part breaks it.\footnote{In this context, another solution is to abandon to have $\sin\theta_{13}=0$ in ${\rm Re}(M^{\mu-\tau}_{\nu})$, which is realized by $M_{e\tau} = \sigma M_{e\mu}$ instead of $M_{e\tau} = -\sigma M_{e\mu}$ in Eq.(\ref{Eq:Texture23-direct}), and CP violation ceases to be maximal \cite{MaxCPmass}. To discuss $\mu$-$\tau$ symmetry in this type of texture is out of the present scope.}  The mixing angles of $\theta_{12,13}$ are given by
%%%%%%%%%%%%%%%%%%%%
\begin{eqnarray}
\tan 2\theta _{12}  &\approx& 2\sqrt 2 \frac{{\rm Re} \left( M_{e\mu} \right)}{{\rm Re} \left( M_{\mu\mu} \right) - \sigma {\rm Re} \left( M_{\mu\tau}\right) - {\rm Re} \left( M_{ee}\right)},
\nonumber\\
\tan 2\theta _{13} e^{i\delta } &=& 2\sqrt 2 \sigma \frac{i{\rm Im} \left( M_{e\mu} \right)}{{\rm Re}\left( M_{\mu\mu} \right) + \sigma {\rm Re}\left( M_{\mu\tau}\right) + {\rm Re}\left( M_{ee}\right)},
\label{Eq:ApproxAnglesSummary-1}
\end{eqnarray}
%%%%%%%%%%%%%%%%%%%%
from Eqs.(\ref{Eq:Angle12-1}) and (\ref{Eq:Angle13-1}).  The expression of $\tan 2\theta _{12}$ is obtained by taking the approximation $\sin^2\theta_{13}\approx 0$.  The maximal CP violation by $e^{i\delta}=\pm i$ is explicitly obtained. 

Summarizing our discussions, we have advocated to use the possibility that the real part of $M_\nu$ only respects the $\mu$-$\tau$ symmetry.  This possibility is extended to the more general case of $M_\nu=M_{+\nu}+M_{-\nu}$ in Eq.(\ref{Eq:Texture23-complex}), where $M_{+\nu}$ serves as a $\mu$-$\tau$ symmetric texture and the symmetry-breaking term of $M_{-\nu}$ acts as a source of $\sin\theta_{13}\neq 0$.  The consistency of the texture is given by the property that particular combinations of $z$, $z^\ast$ and $e^{i\theta}$ become real or pure imaginary.  This property ensures the appearance of real values of $\theta_{12,13}$ while the real value of $\theta_{23}$ arises from $\tan \theta _{23}  = {\rm Im}( {\rm\bf M}_{e\mu})/{\rm Im}({\rm\bf M}_{e\tau})$.  It should be noted that $\theta _{23}$ is not determined by $\tan\theta_{23}=-{\rm Re}\left({\rm\bf M}_{e\tau}\right)/{\rm Re}\left({\rm\bf M}_{e\mu}\right)$ as in the $\mu$-$\tau$ symmetric texture because the Dirac CP violation phase is now active.  It turns out that $M_\nu=e^{i\theta/2}[{\rm Re}(e^{-i\theta/2} M_\nu) +i{\rm Im}(e^{-i\theta/2} M_\nu) ]$, which gives no intrinsic Majorana CP violation while the Dirac CP violation becomes maximal.

%\appendix
%%-------------------------------------------------
%% Appendix
%%-------------------------------------------------
%\section{\label{sec:Appendix}Neutrino Masses and Mixing Angles}

%\input appendix.tex

%%-------------------------------------------------
%% References
%%-------------------------------------------------
%%\begin{references}
%%\end{references}

%%%%%%%%%%%%%%%%%%%%%%%%%%%%%%%%%%%%%%%%%%%%%%%%%%%%%%%%%%%%%%%%%%%%%%%%%%%%%%%%%%%%%%%%


\begin{thebibliography}{}
%%-------------------------------------------------
%% References
%%-------------------------------------------------
%%%%%%%%%%%%%%%%%%%%%%%%%%%%%%%%%%%%%%%%%%%%%%
\bibitem{SK}
	Y. Fukuda {\it et al.}. [Super-Kamiokande Collaboration], \Journal{\PRL}{81}{1158}{1998};
	[\Journal{\Erratum}{81}{4297}{1998}]; \Journal{\PRL}{82}{2430}{1999}.
	See also,
	T. Kajita and Y. Totsuka, \Journal{\RMP}{73}{85}{2001}.
%%%%%%%%%%
\bibitem{experiments}
	Q.A. Ahmed. {\it et al.}, [SNO Collaboration], \Journal{\PRL}{87}{071301}{2001}; \Journal{\PRL}{89}{011301}{2002};
	S. H. Ahn, {\it et al.}, [K2K Collaboration], \Journal{\PLB}{511}{178}{2001}; \Journal{\PRL}{90}{041801}{2003};
	K. Eguchi, {\it et al.}, [KamLAND collaboration], \Journal{\PRL}{90}{021802}{2003};
	M. Apollonio, {\it et al.}, [CHOOZ Collaboration], \Journal{\EPJ}{27}{331}{2003}.
%%%%%%%%%%
\bibitem{NeutrinoSummary}
See for example, 
	R.N. Mohapatra, {\it et al.},``Theory of Neutrinos", [arXive:hep-ph/0412099].
See also,
	S. Goswami, Talk given at {\it Neutrino 2004: The 21st International Conference on Neutrino Physics and Astrophysics}, Paris, France (June 14-19, 2004);
	G. Altarelli, Talk given at {\it Neutrino 2004: The 21st International Conference on Neutrino Physics and Astrophysics}, Paris, France (June 14-19, 2004);
	A. Bandyopadhyay, \Journal{\PLB}{608}{115}{2005}.
%%%%%%%%%%
\bibitem{Nishiura}
	T. Fukuyama and H. Nishiura, in {\it Proceedings of International Workshop on Masses and Mixings of Quarks and Leptons} edited by Y. Koide (World Scientific, Singapore, 1997), p.252; ``Mass Matrix of Majorana Neutrinos", [arXive:hep-ph/9702253];
	Y. Koide, H. Nishiura, K. Matsuda, T. Kikuchi and T. Fukuyama, \Journal{\PRD}{66}{093006}{2002};
	Y. Koide, \Journal{\PRD}{69}{093001}{2004};
	K. Matsuda and H. Nishiura, \Journal{\PRD}{69}{117302}{2004}; \Journal{\PRD}{71}{073001}{2005}.
%%%%%%%%%%
\bibitem{mu-tau}
	C.S. Lam, \Journal{\PLB}{507}{214}{2001}; \Journal{\PRD}{71}{093001}{2005};
	E. Ma and M. Raidal, \Journal{\PRL}{87}{011802}{2001}; [\Journal{\Erratum}{87}{159901}{2001}];
	T. Kitabayashi and M. Yasu\`{e}, \Journal{\PLB}{524}{308}{2002}; 
	\Journal{\IJMP}{17}{2519}{2002}; \Journal{\PRD}{67}{015006}{2003};
	P.F. Harrison and W.G. Scott, \Journal{\PLB}{547}{219}{2002}; 
	E. Ma, \Journal{\PRD}{66}{117301}{2002};
	I. Aizawa, M. Ishiguro, T. Kitabayashi and M. Yasu\`{e}, \Journal{\PRD}{70}{015011}{2004}; I. Aizawa, T. Kitabayashi and M. Yasu\`{e}, \Journal{\PRD}{71}{075011}{2005}.
%%%%%%%%%%
\bibitem{mu-tau1}
	W. Grimus and L. Lavoura,  \Journal{\JHEP}{0107}{045}{2001}; \Journal{\EPJ}{28}{123}{2003}; \Journal{\PLB}{572}{189}{2003}; \Journal{\JPG}{30}{1073}{2004};
``S3$\times$Z2 Model for Neutrino Mass Matrices", [arXive:hep-ph/0504153];
	W. Grimus, A.S. Joshipura, S. Kaneko, L. Lavoura, H. Sawanaka and M. Tanimoto, \Journal{\JHEP}{0407}{078}{2004}; \Journal{\NPB}{713}{151}{2005}; M. Tanimoto, ``Prediction of $U_{e3}$ and $\cos 2\theta_{23}$ from Discrete Symmetry", [arXive:hep-ph/0505031].
%%%%%%%%%%
\bibitem{mu-tau2}
	R.N. Mohapatra,  \Journal{\JHEP}{0410}{027}{2004};
	R.N. Mohapatra and S. Nasri,  \Journal{\PRD}{71}{033001}{2005};
	R.N. Mohapatra, S. Nasri and H. Yu, \Journal{\PLB}{615}{231}{2005}.
%%%%%%%%%%
\bibitem{CPViolation}
	For a recent review, O. Mena, \Journal{\MPL}{20}{1}{2005}.
See also,
	J. Burguet-Castell, M.B. Gavela, J.J. Gomez-Cadenas, P. Hernandez and O. Mena, \Journal{\NPB}{646}{301}{2002};
	W.J. Marciano, ``Extra Long Baseline Neutrino Oscillations and CP Violation", [arXive:hep-ph/0108181];
	J. Burguet-Castell and O. Mena, \Journal{\NIMA}{503}{199}{2003};
	T. Ota and J. Sato, \Journal{\PRD}{67}{053003}{2003};
	T. Ota, \Journal{\JPG}{29}{1869}{2003}; 
	M.V. Diwan, \Journal{\IJMP}{18}{4039}{2003};
	H. Minakata and H. Sugiyama, \Journal{\PLB}{580}{216}{2004};
	O. Mena and S. Parke, \Journal{\PRD}{70}{093011}{2004};
    M. Ishitsuka, T. Kajita, H. Minakata and H. Nunoka, ``Resolving Neutrino Mass Hierarchy and CP Degeneracy by Two Identical Detectors with Different Baselines", [arXive:hep-ph/0504026].
%%%%%%%%%%%%%%%%%%%%%%%%%%%%%%%%%%%%%%%%%%%%%%
\bibitem{PMNS} 
	B. Pontecorvo, \Journal{\JETPUSSR}{7}{172}{1958} [\Journal{\ZETP}{34}{247}{1958}];
	Z. Maki, M. Nakagawa and S. Sakata, \Journal{\PTP}{28}{870}{1962}. 
%%%%%%%%%%%%%%%%%%%%%%%%%%%%%%%%%%%%%%%%%%%%%%
\bibitem{CPphases} 
	S.M. Bilenky, J. Hosek and S.T. Petcov, \Journal{\PLBOLD}{94B}{495}{1980};
	J. Schechter and J.W.F. Valle, \Journal{\PRD}{22}{2227}{1980};
	M. Doi, T. Kotani, H. Nishiura, K. Okuda and E. Takasugi, \Journal{\PLBOLD}{102B}{323}{1981}.
%%%%%%%%%%
\bibitem{CPinMixing} 
See for example,
S.T. Petcov. \Journal{\NPSUPPL}{143}{159}{2005}.
%%%%%%%%%%
\bibitem{MassTextureCP1}
	See for example,
	M. Frigerio and A.Yu. Smirnov, \Journal{\NPB}{640}{233}{2002}; \Journal{\PRD}{67}{013007}{2003};
	S.F. King, in {\it Proceedings of 10th International Workshop on Neutrino Telescopes} edited by  M. Baldo-Ceolin (U. of Padua Publication, Italy, 2003), ``Neutrino Mass, Flavor and CP Violation", [arXive:hep-ph/0306095];
	Z.Z. Xing, \Journal{\IJMP}{19}{1}{2004};
	O.L.G. Peres and A.Yu. Smirnov, \Journal{\NPB}{680}{479}{2004};
	C.H. Albright, \Journal{\PLB}{599}{285}{2004};
	J. Ferrandis and S. Pakvasa, \Journal{\PLB}{603}{184}{2004};
	S. Zhou and Z.Z. Xing, \Journal{\EPJ}{38}{495}{2005};
	S.T. Petcov and W. Rodejohann, \Journal{\PRD}{71}{073002}{2005};
	G.C. Branco and M.N. Rebelo, \Journal{\NJP}{7}{86}{2005};
	S.S. Masood, S. Nasri and J. Schechter, \Journal{\PRD}{71}{093005}{2005}.
%%%%%%%%%%
\bibitem{MassTextureCP2}
	E. Ma, \Journal{\MPL}{17}{2361}{2002};
	K.S. Babu, E. Ma and J.W.F. Valle, \Journal{\PLB}{552}{207}{2003};
	W. Grimus and L. Lavoura, \Journal{\PLB}{579}{113}{2004};
	P.F. Harrison and W.G. Scott, \Journal{\PLB}{594}{324}{2004}.
%%%%%%%%%%
\bibitem{CP2-3}
	R.N. Mohapatra, S. Nasri and H. Yu, in Ref.\cite{mu-tau2}.
%%%%%%%%%%
\bibitem{CP2-3-1}
W. Grimus and L. Lavoura, ``S3$\times$Z2 Model for Neutrino Mass Matrices", [arXive:hep-ph/0504153] in Ref.\cite{mu-tau1}.
%%%%%%%%%%
\bibitem{MaxCPmass} 
	I. Aizawa, T. Kitabayashi and M. Yasu\`{e}, ``Neutrino Mass Textures with Maximal CP Violation", [arXive:hep-ph/0504172].
%%%%%%%%%%
\bibitem{GeneralCP} 
	I. Aizawa and M. Yasu\`{e}, \Journal{\PLB}{607}{267}{2005}.
\end{thebibliography}
\end{document}